\documentclass[12]{article}
\usepackage{amsmath,amssymb,amsfonts,times,graphicx,hyperref}
\begin{document}

\title{New Universal Theory of Injury Prediction and Prevention}
\author{Dr. Vladimir G. Ivancevic \\
{\small Vladimir.Ivancevic@dsto.defence.gov.au}}
\date{}
\maketitle

\noindent The prediction and prevention of traumatic brain injury, spinal
injury and musculo-skeletal injury is a very important aspect of preventive
medical science. Recently, in a series of papers \cite{ivbrain,ivspine,ivgen}%
, I have proposed a new coupled loading-rate hypothesis as a unique cause of
all above injuries. This new hypothesis states that the main cause of all
mechanical injuries is a Euclidean Jolt, which is an impulsive loading that
strikes any part of the human body (head, spine or any bone/joint) -- in
several coupled degrees-of-freedom simultaneously. It never goes in a single
direction only. Also, it is never a static force. It is always an impulsive
translational and/or rotational force coupled to some mass eccentricity.
This is, in a nutshell, my universal Jolt theory of all mechanical injuries.%
\begin{figure}[h]
\centerline{\includegraphics[width=11cm]{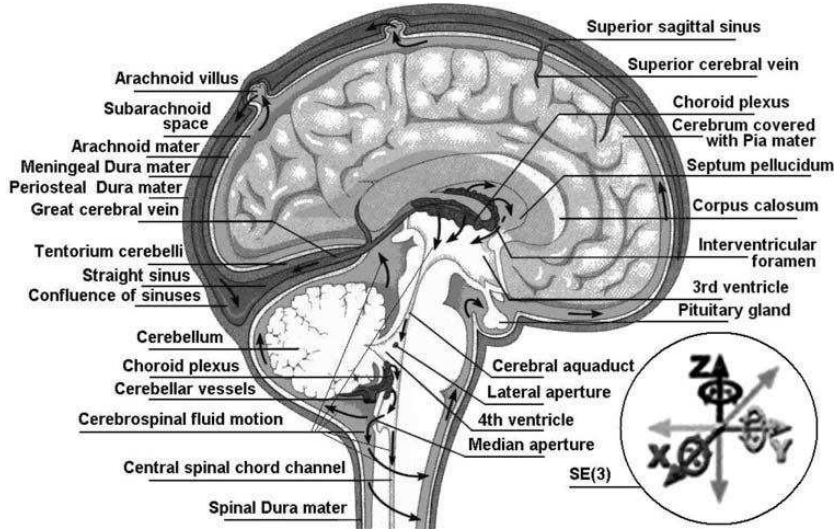}}
\caption{Human brain and its
SE(3)-group of microscopic
three-dimensional motions
within the cerebrospinal fluid
inside the cranial cavity.}
\label{BrainInj}
\end{figure}

To show this, based on the previously defined covariant force law, I have
firstly formulated the fully coupled Newton--Euler dynamics of:

1.\quad Brain's micro-motions within the cerebrospinal fluid inside the
cranial cavity;

2.\quad Any local inter-vertebral motions along the spine; and

3.\quad Any local joint motions in the human musculo-skeletal system.\newline

Then, from it, I have defined the essential concept of \textbf{Euclidean Jolt%
}, which is the main cause of all mechanical injuries. The Euclidean Jolt
has two main components:

1.\quad Sudden motion, caused either by an accidental impact or slightly
distorted human movement; and

2.\quad Unnatural mass distribution of the human body (possibly with some
added masses), which causes some mass eccentricity from the natural
physiological body state.
\begin{figure}[h]
\centerline{\includegraphics[width=11cm]{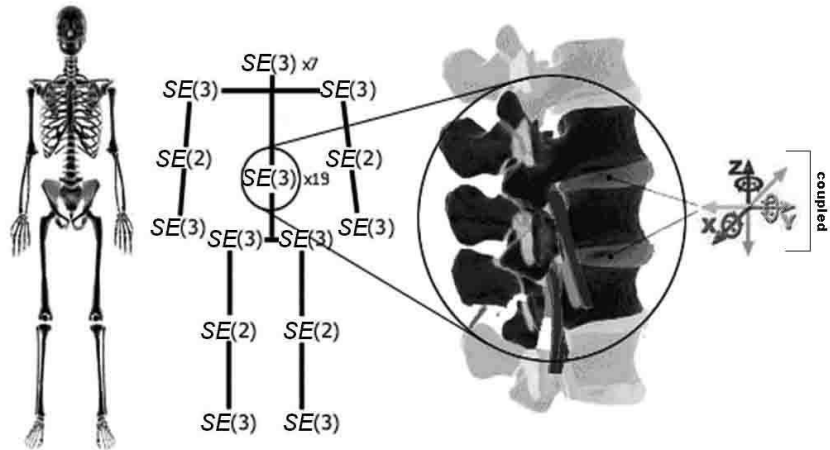}}
\caption{Human body representation in terms of SE(3)/SE(2)-groups of rigid-body motion, with
the vertebral column represented as a chain of 26 flexibly-coupled SE(3)-groups.}
\label{SpineSE3}
\end{figure}

What does this all mean? I'll try to explain it in \textquotedblleft plain
English\textquotedblright. As we live in a 3D space, one could think that
motion of any part of the human body, either caused by an accidental impact
or by voluntary human movement, \textquotedblleft just obeys classical
mechanics in 6 degrees-of-freedom: three translations and three
rotations\textquotedblright . However, these 6 degrees-of-freedom are not
independent motions as it is suggested by the standard term
\textquotedblleft degrees-of-freedom\textquotedblright . In reality, these
six motions of any body in space are coupled. Firstly, three rotations are
coupled in the so-called rotation group (or matrix, or quaternion).
Secondly, three translations are coupled with the rotation group to give the
full Euclidean group of rigid body motions in space. A simple way to see
this is to observe someone throwing an object in the air or hitting a tennis
ball: how far and where it will fly depends not only on the standard
\textquotedblleft projectile\textquotedblright\ mechanics, but also on its
local \textquotedblleft spin\textquotedblright\ around all three axes
simultaneously. Every golf and tennis player knows this simple fact. Once
the spin is properly defined we have a \textquotedblleft fully coupled
Newton--Euler dynamics\textquotedblright\ -- to start with.\newline

The covariant force law for any biodynamical system (which I introduced
earlier in my biodynamics books and papers, see my references in the cited
papers above) goes one step beyond the Newton--Euler dynamics. It states:
\begin{equation*}
\mathbf{Euclidean\ Force\ covector\ field\ =\ Body\ mass\ distribution\
\times \ Euclidean\ Acceleration\ vector\ field}
\end{equation*}

This is a nontrivial biomechanical generalization of the fundamental
Newton's definition of the force acting on a single particle. Unlike
classical engineering mechanics of multi-body systems, this fundamental law
of biomechanics proposes that forces acting on a multi-body system and
causing its motions are fundamentally different physical quantities from the
resulting accelerations. In simple words, forces are massive quantities
while accelerations are massless quantities. More precisely, the
acceleration vector field includes all linear and angular accelerations of
individual body segments. When we couple them all with the total body's
mass-distribution matrix of all body segments (including all masses and
inertia moments), we get the force co-vector field, comprising all the
forces and torques acting on the individual body segments. In this way, we
have defined the 6-dimensional Euclidean force for an arbitrary
biomechanical system.
\begin{figure}[h]
\centerline{\includegraphics[width=7cm]{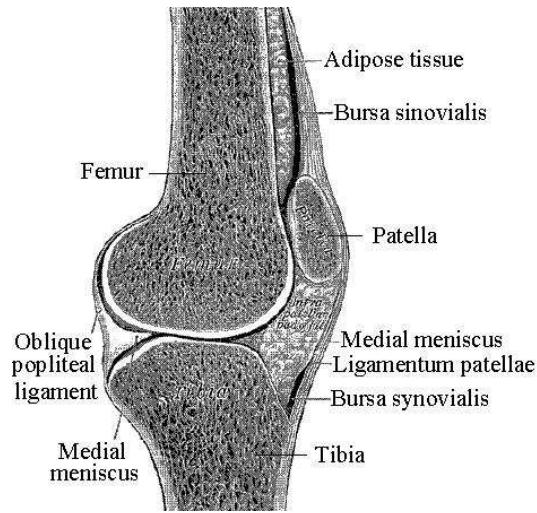}}
\caption{Schematic latero-frontal view of the left knee joint. Although designed to perform mainly
flexion/extension (strictly in the sagittal plane) with some restricted medial/lateral rotation in
the semi-flexed position, it is clear that the knee joint really has at least six-degrees-of-freedom,
including three micro-translations. The injury actually occurs when some of these microscopic
translations become macroscopic, which normally happens only after an external jolt.}
\label{KneeInj}
\end{figure}

Now, for prediction of injuries, we need to take the rate-of-change (or
derivative, with respect to time) of the Euclidean biomechanical force
defined above. In this way, we get the Euclidean Jolt, which is the sudden
change (in time) of the 6-dimensional Euclidean force:
\begin{equation*}
\mathbf{Euclidean\ Jolt\ covector\ field\ =\ Body\ mass\ distribution\
\times \ Euclidean\ Jerk\ vector\ field}
\end{equation*}

And again, it consists of two components: (i) massless linear and angular
jerks (of all included body segments), and (ii) their mass distribution. For
the sake of simplicity, we can say that the mass distribution matrix
includes all involved segmental masses and inertia moments, as well as
\textquotedblleft eccentricities\textquotedblright\ or \textquotedblleft
pathological leverages\textquotedblright\ from the normal physiological
state.\newline

Therefore, the unique cause of all brain, spine and musculo-skeletal
injuries has two components:

1.\quad Coupled linear and angular jerks; ~and

2.\quad Mass distribution with \textquotedblleft
eccentricities\textquotedblright .\newline

\noindent In other words, ~\textbf{there are no injuries in static
conditions without any mass eccentricities; all injuries are caused by
mutually coupled linear and angular jerks, which are also coupled with the
involved human mass distribution.\newline
}
Note the difference between jerk and jolt. For example, sharp braking in a car causes jerk, while actual colliding with another object causes jolt. And it is always in several directions and rotations combined.\\

The Euclidean Jolt causes two forms of discontinuous brain, spine or
musculo-skeletal injury:

1.\quad Mild rotational disclinations; ~and

2.\quad Severe translational dislocations (or, fractures).\newline

In the cited papers above, I have developed the soft-body dynamics of
biomechanical disclinations and dislocations, caused by the Euclidean Jolt,
using the Cosserat multipolar viscoelastic continuum model.

Implications of the new universal theory are various, as follows.\newline

\noindent\textbf{A.} ~~The research in traumatic brain injury (TBI, see Figure \ref{BrainInj}) has so
far identified the rotation of the brain-stem as the main cause of the TBI
due to various crashes/impacts. The contribution of my universal Jolt theory
to the TBI research is the following:\newline

1.\quad Rigorously defined this brain rotation as a mechanical disclination
of the brain-stem tissue modelled by the Cosserat multipolar soft-body model;

2.\quad Showing that brain rotation is never uni-axial but always
three-axial;

3.\quad Showing that brain rotation is always coupled with translational
dislocations. This is a straightforward consequence of my universal Jolt
theory.\newline

These apparently `obvious' facts are actually \textsl{radically new:} we
cannot separately analyze rapid brain's rotations from translations, because
they are in reality always coupled.\newline

One practical application of the brain Jolt theory is in design of helmets.
Briefly, a `hard' helmet saves the skull but not the brain; alternatively, a
`soft' helmet protects the brain from the collision jolt but does not
protect the skull. A good helmet is both `hard' and `soft'. A proper helmet
would need to have both a hard external shell (to protect the skull) and a
soft internal part (that will dissipate the energy from the collision jolt
by its own destruction, in the same way as a car saves its passengers from
the collision jolt by its own destruction).\newline

Similarly, in designing safer car air-bags, the two critical points will be
(i) their placement within the car, and (ii) their \textquotedblleft
soft-hard characteristics\textquotedblright , similar to the helmet
characteristics described above.\newline

\noindent\textbf{B.} ~~In case of spinal injury (see Figure \ref{SpineSE3}), the contribution of my
universal Jolt theory is the following:\newline

1.\quad The spinal injury is always localized at the certain vertebral or
inter-vertebral point;

2.\quad In case of severe translational injuries (vertebral fractures or
discus herniae) they can be identified using X-ray or other medical imaging
scans; in case of microscopic rotational injuries (causing the back-pain
syndrome) they cannot be identified using current medical imaging scans;

3.\quad There is no spinal injury without one of the following two causes:

\qquad a.~~~Impulsive rotational + translational loading caused by either
fast human movements or various\newline
crashes/impacts; and/or

\qquad b.~~~Static eccentricity from the normal physiological spinal form,
caused by external loading;

\qquad c.~~~Any spinal injury is caused by a combination of the two points
above: impulsive rotational +\newline
translational loading and static eccentricity.\newline

This is a straightforward consequence of my universal Jolt theory. We cannot
separately analyze translational and rotational spinal injuries. Also, there
are no \textquotedblleft static injuries\textquotedblright\ without
eccentricity. Indian women have for centuries carried bulky loads on their
heads without any spinal injuries; they just prevented any load
eccentricities and any jerks in their motion.\newline

The currently used \textquotedblleft Principal loading
hypothesis\textquotedblright\ that describes spinal injuries in terms of
spinal tension, compression, bending, and shear, covers only a small subset
of all spinal injuries covered by my universal Jolt theory. To prevent
spinal injuries we need to develop spinal jolt awareness: ability to control
all possible impulsive spinal loadings as well as static eccentricities.%
\newline

\noindent\textbf{C.} ~~In case of general musculo-skeletal injury (see Figure \ref{KneeInj} for the particular case of knee injury), the
contribution of my universal Jolt theory is the following:\newline

1.\quad The injury is always localized at the certain joint or bone and
caused by an impulsive loading, which hits this particular joint/bone in
several coupled degrees-of-freedom simultaneously;

2.\quad Injury happens when most of the body mass is hanging on that joint;
for example, in case of a knee injury, when most of the body mass is on one
leg with a semi-flexed knee --- and then, caused by some external shock, the
knee suddenly \textquotedblleft jerks\textquotedblright\ (this can happen in
running, skiing, and ball games, as well as various crashes/impacts); or, in
case of shoulder injury, when most of the body mass is hanging on one arm
and then it suddenly jerks.\newline

To prevent these injuries we need to develop musculo-skeletal jolt
awareness. For example, never overload a flexed knee and avoid any kind of
uncontrolled motions (like slipping) or collisions with external objects.

\end{document}